# Semantics of UML 2.0 Activity Diagram for Business Modeling by Means of Virtual Machine

Valdis Vitolins, Audris Kalnins

*Abstract*— The paper proposes a more formalized definition of UML 2.0 Activity Diagram semantics. A subset of activity diagram constructs relevant for business process modeling is considered. The semantics definition is based on the original token flow methodology, but a more constructive approach is used. The Activity Diagram Virtual machine is defined by means of a metamodel, with operations defined by a mix of pseudocode and OCL pre- and postconditions. A formal procedure is described which builds the virtual machine for any activity diagram. The relatively complicated original token movement rules in control nodes and edges are combined into paths from an action to action. A new approach is the use of different (push and pull) engines, which move tokens along the paths. Pull engines are used for paths containing join nodes, where the movement of several tokens must be coordinated. The proposed virtual machine approach makes the activity semantics definition more transparent where the token movement can be easily traced. However, the main benefit of the approach is the possibility to use the defined virtual machine as a basis for UML activity diagram based workflow or simulation engine.

## I. INTRODUCTION

The UML 2.0 standard (Standard), whose development was started in 2002, now is under the final adoption [1]. Activity diagrams (AD) are redesigned radically in the Standard, where besides significant syntax modification the main difference is switching from State Machine based semantics to the token flow (Petri net like) semantics. The semantics sections in the Standard contain an informal description for each AD element, how this element influences the token movement in a diagram. Though these descriptions of AD semantics are sufficient for informal modeling of processes, the formality level is not sufficient for the use of activity diagrams as precise process definitions, e.g., for workflow specification.

The goal of this paper is to define an Activity Diagram Virtual Machine (ADVM), which would describe this token-based semantics formally enough for process execution. The paper proposes an ADVM for a subset of activity diagram features, which are significant for business process modeling and definition [2]. ADVM itself is defined by means of a runtime extension to the metamodel, whose operations are defined by means of a procedural OCL-based pseudocode and pure OCL pre- and postconditions [3]. Though the size of the paper does not permit to present completely formal descriptions for all operations of ADVM, the provided methodology is sufficient for achieving this goal. The structure of the proposed ADVM is, on the one hand, meant to be simple enough for precise analysis of process definitions by humans. On the other hand, it could be used as the basis, e.g., for an activity diagram based workflow engine. It should be noted that a similar approach could be used also for "cleaning up" all the remaining semantic problems in the Standard, e.g., those related to single execution.

The definition of ADVM in the paper is divided into two parts. Section IV shows how a "runtime copy" is built for an activity diagram. Section V provides the definitions of all runtime operations. A possible usage of the approach is given in the conclusion.

There are only few papers commenting the intended informal semantics of AD [2,8,9], even less research is devoted to formal definition of UML 2.0 activity diagram semantics. Many of them use pure Petri nets, in such a way losing some semantics for data tokens [4,5]. The approach closest to the one used in this paper is in [6]. There a similar subset including object flows is analyzed. The main difference is that [6] uses a translator to classic (colored) Petri nets. Though this enables the use of formal process analyzers, Petri nets are not the best model for understanding workflows [4]. Therefore our approach which tries to rely on the original AD notation as much as possible in semantics definition is more suited for workflows, even if it doesn't support a formal mathematical analysis.

## II. SUBSET OF THE UML 2.0 ACTIVITY DIAGRAM AND LIMITATIONS

### A. Used Subset of the UML Activity Diagram

Since the number of concepts and the metamodel for UML 2.0 activity diagrams is very large, only "the most essential" subset of AD is considered in this paper. On the other hand, focus is on those AD elements, which directly influence the token movement semantics. Finally, a number of useful constructs are not included just to limit the paper size. But the main idea is to show that the used techniques work for the precise formalization of a subset of AD reasonable for process modeling. More concepts for which these techniques should work also are listed in the conclusion.

Manuscript received in April 1, 2005. This work was supported in part by European Social Fund (ESF).

Valdis Vitolins, University of Latvia, IMCS, 29 Raina blvd, Riga, Latvia, Valdis_Vitolins@exigengroup.lv

Audris Kalnins, University of Latvia, IMCS, 29 Raina blvd, Riga, Latvia, Audris.Kalnins@mii.lu.lv

In the Standard the activity diagram definition is structured into packages with ever growing complexity – Fundamental, Basic, Intermediate, Complete (and some other which form a sideline of no great interest for business modeling). The subset analyzed in this paper contains (almost completely) concepts from Fundamental to Intermediate activities. We have included Activities, Actions, Control and Object flows, Activity Parameter nodes, Pins and all Control nodes. Namely these concepts are most relevant for the semantics definition by means of token movement. From Intermediate Activities, we have not included Activity Partitions and Groups, because they are not relevant for token management.

Now more detailed comments on the subset. The most essential is the choice of subclasses of Object nodes. Only the Pins (Input and Output) and Activity Parameter nodes are included. Central buffers (with relatively complicated semantics) are not included because they are mostly used for physical system modeling [8]. No explicit object nodes are used for object flows – only the output and input pins at ends. Certainly, there is a certain controversy about object nodes in the Standard itself – ObjectNode is an abstract class, but has a graphical notation and a defined semantics.

There are also restrictions on flows. First, object flows must mandatory have pins at points, where they are connected to actions (certainly, not at control nodes, exceptions see below). On the one hand, the existence of explicit pins makes the semantics definition much more transparent and similar to Petri nets – there are places for tokens to live in. On the other hand, this notation has already been accepted in practice – the most advanced UML 2.0 tool at this moment (IBM Rational's RSA [7]) uses only this notation. In addition, there is a natural restriction that each flow leaving (or entering) an action has a separate pin. Pins can have no upper bounds, and they have either a specified type or no type – for accepting values of several different types (e.g., the result of a Join). In the current version of Standard it is permitted to have also control pins (i.e., pins with *isControlType*=true). Using this fact, we require in our subset that control flows also must have pins at any (possible) end. Just to make the description of our ADVM shorter, we use in this paper a different criterion: control pins (and tokens) have the NULL type. Thus, in fact we can consider control flows to be a special case of object flows. Certainly, all the semantics specific only to control flows and tokens is retained. The final assumption on flows is that initial nodes also must have output (control) pins and final nodes must have input pins (of any relevant type). Though not suggested by the Standard, it does not contradict either (e.g., it is asserted, that a CentralBufferNode might be used after the initial node), and the semantics is not changed this way. Thus, flows have pins wherever possible. Fig. 2 shows a diagram example in our subset. Actually, all this section on flows just restricts the diagram drawing – any reasonable Intermediate level diagram can be redrawn this way.

The next issue is actions. From all the action types available at Intermediate level only the CallBehaviorAction makes sense for business modeling, therefore only this was included in the subset [9]. This action can invoke either an OpaqueBehavior (elementary task) or another Activity.

An additional requirement is that at any Decision node the outgoing edges must have mutually exclusive guards – a requirement found useful also in the Standard. The main real restriction for decisions (and any edge in general) is that guards are not allowed to use data from the activity context (i.e., only the data from the current token may be used).

Only one element from Complete activities is included – join specifications (they also may use token data only).

The selected subset implies two general restrictions on token movement – there may be no real "race for token" by several actions, and the guard value on a token cannot change in time. See more on this in section VI.

The Intermediate level uses separate execution, so only this mode is considered in the paper. This mode is sufficient for most cases of process modeling, e.g., workflow definition. Though there are some semantic problems with single execution in the Standard, our approach could be extended to single execution quite easily (see IV.A).

Fig. 3 shows the metamodel of the selected subset (light classes). To reduce the number of classes, the package merge has been performed, and some unnecessary for the paper classes and inheritance hierarchies have also been removed.

Capitalized names are used in the paper as exact references to the metamodel classes (e.g. Activity), lowercase names are used as a generic term or arbitrary instance of this class (e.g. token).

*B. Additional restrictions for Activity Diagrams*

In addition to the restrictions imposed by UML 2.0, we assume that Activity Diagram is correct if and only if:
a) outgoing edge from a ControlNode is not an incoming edge for the same ControlNode
b) there should be no paths between CallBehaviorActions, InitialNodes, FinalNodes or ActivityParameterNodes containing both ForkNodes and JoinNodes.

So, invalid cases for a) and b) are:

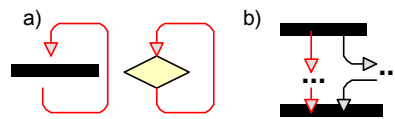

**Fig. 1** Invalid constructs for activity diagram – loops and ForkNodes in the same path with JoinNodes

Reason for a) is obvious - it prevents deadlocks for control nodes waiting for input, which should be provided by themselves.

Restriction b) is reasonable from the practical point of view, because there is no need to make parallel branches, if they are simply joined back without any operation in these branches (i.e. there is no CallBehaviorAction between them). This restriction significantly simplifies rules for token movement.

We assume that AD is validated before the diagram execution.

III. GENERAL DESCRIPTION OF UML 2.0 AD AND OUR VM

Let us give an example of activity diagram in the described subset. This example is similar to the one used in the Standard, but with some modifications illustrating all the key elements. Actually the process is described by two AD, and the main AD *Process Order* invokes another one – *Make Payment*. The main diagram starts with the initial node, then the process flows through decision-, fork-, join- and merge nodes and finishes in ActivityFinal node. The *MakePayment* action invokes the subordinated activity, which starts and finishes with activity parameter nodes. All flows (control and object) have pins at their action ends (and also at initial and final nodes) – as it is required by our subset. Control pins have no type. Note that an object pin also may be typeless (the input pin of *Close Order*), here this is the only way to preserve the type conformance after merge and join nodes. Namely, this input pin must be able to hold tokens of types both *Order* and *Payment*. The advantage for semantics definition of using all these explicit pins is that we always have a place for tokens to "live in" – much in the same way as places in Petri nets are used.

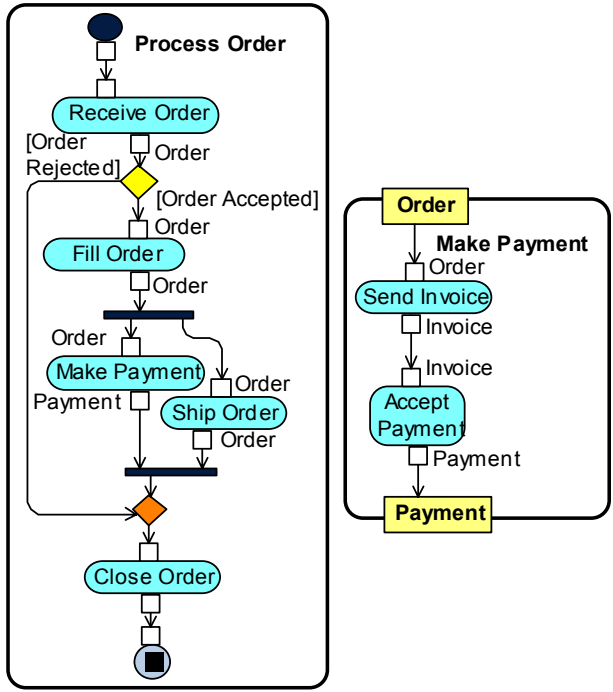

**Fig. 2** Example activity diagram "Process order" which invokes sub activity diagram "Make Payment" from action "Make Payment"

*A. Standard semantics of Activity diagrams*

In Standard the activity diagram semantics is described in a highly distributed manner, where each AD element has its role in AD execution - fork, decision, merge and join nodes each process token flow in their own way. E.g., for a ForkNode, "when an offered token is accepted on all the outgoing edges, duplicates of the token are made and one copy traverses each edge". But for JoinNode, "if there is a token offered on all incoming edges, then tokens are offered on the outgoing edge". The term **offered** actually needs to be understood more formally. The sole reasonable interpretation of "**offered**" is that a token is not actually moved along the edge, but only becomes visible through this edge. When there is a sequence of edges and control nodes in a diagram, these "offering visibility" rules define a sort of "transitive closure visibility", by which tokens from output pins become visible to their actual consumers – actions, certainly, via their input pins.

Now, according to the Standard, an Action is executed "when all of the input pins are offered tokens and accept them all at once, precluding them from being consumed by any other actions". Namely at this moment the required set of tokens move to their corresponding input pins. This actually means that action uses **pull semantics** for token processing – the only really active elements in a diagram are the "action engines", which try to fill up their input pins with fresh sets of tokens, to be consumed by these actions. The "all at once" phrase in the definition actually means that all tokens from each output pin, which "offers" tokens (is visible) to the action are consumed. Especially, if there is a join before the action, which joins object flows, then all tokens from these output pins are "serialized" and provided to the relevant input pin of the action as one coherent group. It should be noted that this semiformal semantics is well defined only for separate execution (for single execution e.g. the "merging" of control tokens could lead to a loss of concurrent control threads), but this is a topic of a separate paper.

It is clear that a "standard Activity Diagram Virtual Machine" (ADVM) could be defined, with "action engines" as the only active elements. However, the formalization of the entire "offering" (visibility) rules by a sort of "traffic switches" affecting the token movement would be highly complicated – visibility rules are harder to implement than simple actions.

*B. General principles of proposed ADVM*

In this paper we propose a different approach to building an ADVM. Subgraphs of edges and control nodes connecting "stable places" – output and input pins are "truncated" into explicit **paths** leading directly from output pins to input pins in our definition of ADVM. Each path has a **condition** – the guards of its edges "anded" together. Pins in turn may be serviced by active elements – token engines. We introduce two different kinds - **Push** and **Pull** token **engines**.

The same way, there are also two kinds of paths – push and pull paths. **Push paths** are those containing only Decision, Merge and Fork nodes (or no control nodes at all). A push path is "serviced" by a Push engine in its start node – the corresponding output pin. In our subset tokens from an output pin can be pushed via push paths independently from each other directly to their destinations – input pins, whenever path conditions permit it (see a formal justification for it in section VI). Thus token movement is very transparent in the push case.

**Pull paths** are those containing at least one Join node (and Decisions and Merges, but no Forks in our subset!). Pull paths are serviced by a Pull engine at their destination – an input pin. According to the AD semantics, the movement of tokens along pull paths having a common destination must be coordinated – only an adequate set of tokens can jointly pass a Join node. When these tokens are object tokens, then according to the Standard they must jointly continue their travel. Therefore the concept of **Token Group** is introduced in our ADVM – it is the set of tokens, which is jointly pulled by a Pull Engine into its serviced input pin. It should be noted that the same input pin may contain groups with different structure – in case when pull paths contain merge nodes. In, addition, the pulled in groups must satisfy the **join criteria** – "anded" join specifications. For pure control tokens there are no groups – they are "collapsed" into one token according to the Standard semantics.

Thus, the pull engine is much more complicated than the push one – but such is the UML semantics. Pull engine is described in detail in section V.D.

The **Action engine** is much simpler than its counterpart in the original semantics. Its sole task is to seize one token from each input pin (or whole one group, if this is a pull pin), when a complete set is present and to "consume" this set. Certainly, when the action invokes a subordinated activity it has to provide its input parameters and collect the output parameters. The main semantic difference between our and the standard action engine is that for our engine tokens (or groups) are moved (by token engines) independently to each input pin, while (as it was cited) the standard engine pulls them "all at once". However, this cannot lead to serious differences in behavior, since real "races for tokens" by several actions are impossible in our subset – see more in section VI.

Finally, the invocation, start up and termination of an activity are managed by the relatively straightforward **Activity** engine (*ActivityR* class).

The rest of the paper is devoted to the formal description of the proposed ADVM, while the section VI provides a semiformal justification that the semantics formalized by the proposed ADVM is the same standard one (for the selected subset) – the action traces actually coincide.

Thus, the goal of this research has been to provide a complete executable formalization of activity diagram semantics by an ADVM, which could both be analyzed theoretically and serve as a "prototype" for real AD execution (e.g., as a workflow engine). Authors hope that the provided ADVM is more usable for various kinds of formal analysis than the informal original semantics description.

### C. Metamodel extensions and model mapping

In order to define an ADVM formally, the metamodel of AD must be extended. One solution is to add operations to the original metaclasses, but since our ADVM requires new concepts, a more natural solution is to build a special AD runtime metamodel containing appropriate classes with operations. Most of the new classes are "dynamic" counterparts of the corresponding "static" classes of AD metamodel. Fig.3 shows both metamodels combined - the original classes are light and the new ones dark. We remind that the AD metamodel is "flattened" with respect to the standard one in order to reduce its size. Whenever possible, the corresponding classes in both metamodels are linked by special bi-directional associations (so called **mapping associations**, dashed lines). Only the main "internal" associations for both metamodels are depicted in Fig.3.

For figure simplicity, compositions in Fig. 3 and Fig. 4 are drawn as trees, by merging the composition ends into a single segment, according to the Standard presentation options. Adornments on that single segment apply to all of the composition ends. Lines are joined only by "T" junctions, "X" junctions are simple line crosses.

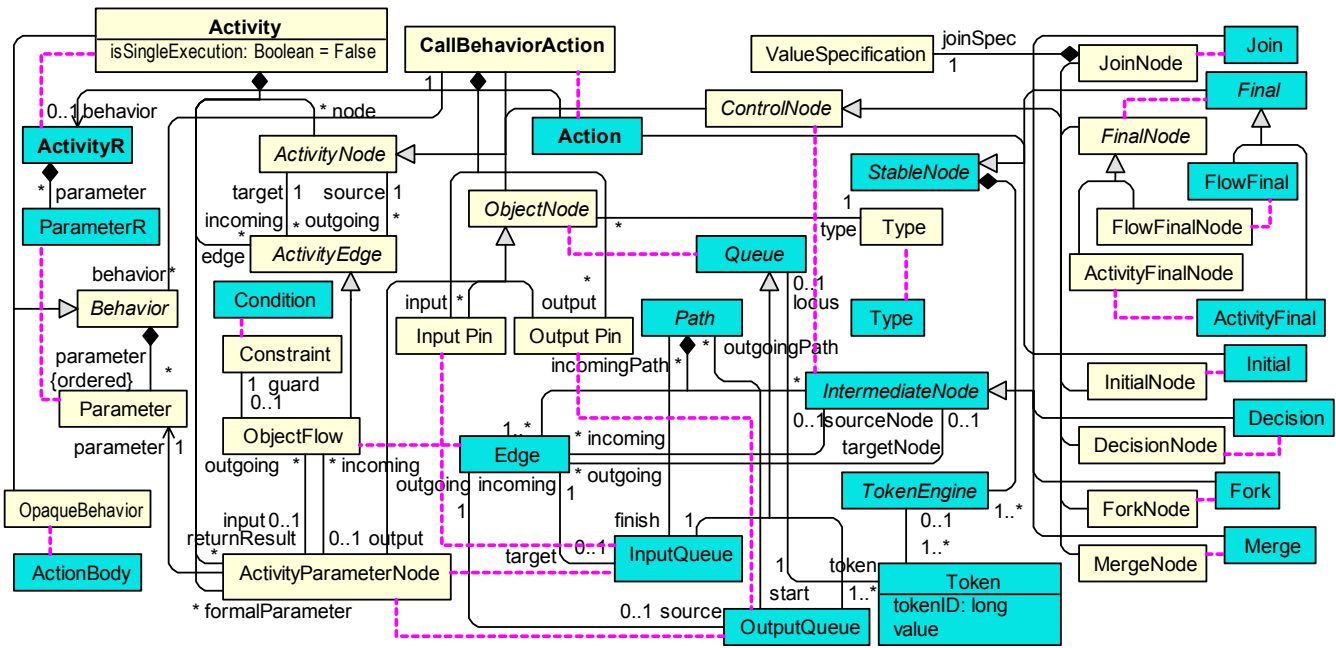

**Fig. 3** Subset of the UML activity diagram and relations to their runtime classes

The main idea is that when an activity is invoked, the corresponding runtime class instances are created for the activity instance and all its components. Namely these instances act as the virtual machine executing the given activity. The creation of these runtime instances is singled out as a separate step in execution and described in section IV.

Actually, the creation of runtime instances from activity model instances is a model transformation, where the source model is the Activity diagram definition, but the target model represents instances of the runtime classes. In that sense Fig.3 represents a "general transformation schema", where the mapping associations have a formal semantics in this transformation. In the direction from a definition class to runtime class it means, that in the transformation process for each instance of the definition class one instance of runtime class should be created. In the opposite direction it shows, from which definition instance the runtime instance is created. This information is used when new instances should be created with specific properties, which can be got only from the source model.

To save space, multiplicities and role names are not shown for mapping associations, but they are 1 at the definition end and 0..1 at runtime end. Such associations are typical for model transformations. Formally we can say that these additional associations are coming from our transformation package, which is merged with a subset of the Standard Activity Diagram metamodel.

In this paper the transformations are described by a mix of pseudocode and OCL postconditions. But they could be described as well by means of specialized transformation languages, such as MOLA [10] or QVT-Merge [11].

### IV. ACTIVITY CONSTRUCTION

#### A. Activity Execution and Invocation

According to the Standard there are two stages of activity performing – creation of the activity execution and activity invocation. Our ADVM assigns a precise meaning to them:

1. Activity execution means creation of the activity runtime instance (ActivityR and its components) and its activation. In this stage only the necessary elements are created, but no tokens or parameters are passed. These operations are done using the ActivityFactory class and ActivityR.*activate()* operation.
2. Activity invocation means activity starting using the *invoke(Object[*])* operation of the ActivityR class. It could be invoked in the following ways:
    a) if the activity has an InitialNode, it is invoked without parameters. Then *invoke()* method puts control tokens into OutputQuee for all InitialNodes.
    b) if the activity has input parameters, it is started using parameters by *invoke(Object[*])*, which places data tokens with appropriate values into all inputParameter nodes (OutputQueues) of the activity.

When the construction of an AD runtime instance is complete, it is activated using the *activate()* method. This method fires processes for token engines, actions, final nodes and activity, and they are ready for token processing, but no tokens are created in this method. Activity invocation only provides a new set of tokens inside an existing activity runtime instance.

For activity creation we use a special ActivityFactory class (Fig. 4). There is only one instance of the ActivityFactory and it is used for creation of each new activity. Thus ActivityFactory plays the role of an entry point for ADVM.

The operation *createActivity(Activity)* of the ActivityFactory is used for construction of the runtime instances. This operation checks, whether the activity has single or separate execution mode. If the execution mode is separate, a new activity runtime instance is always created. If the mode is single, the operation checks, whether an appropriate instance for this Activity already exists and creates a new instance only if it doesn't exist. The *createActivity()* operation returns a reference to the runtime (ActivityR) instance and further management of the activity execution is made through ActivityR operations.

To clarify the sequence of ADVM construction, the creation process is separated into several steps, which are performed through gradual invocation of the factory *create..()* methods: *createAction()*, *creteInputQueue()*, *creteOutputQueue()*, *createIntermediateNode()* methods create runtime instances for the main elements of Activity – Actions, Pins and flow control nodes. *createInitial()*, *createFlowFinal()* and *createActivityFinal()* methods create runtime instances for these kinds of nodes and input/output queues for them.

Using the formalParameter and returnResult associations, activity parameter nodes are fetched and *createInput/OutputQueue()* methods create appropriate input/output queues and runtime parameters (ParameterR) from them. The *createEdges()* method creates runtime instance for each ObjectFlow (using Activity.edge). ActivityEdge Guards in an AD model are transformed into Conditions for Edges.

Then the "nontrivial elements" - paths and token engines are created using *createPaths()*, *createTokenEngines()* and *createJoinCriteria()* methods. These methods will be explained in details in section IV.D. If an activity has no ActivityFinal nodes, a boolean tag is set, which enables a process for checking whether all outputParameter nodes have tokens (an alternative way of completing the execution).

It should be noted, that only the main class instances are created using the factory class. Associated attributes and simple instances are created using local *addNew()* methods.

```
ActivityFactory
createActivity(Activity): ActivityR
createAction(CallBehaviorAction)
createIntermediateNode(ControlNode)
createEdge(ActivityEdge)
createInputQueue(ActivityNode)
createOutputQueue(ActivityNode)
createInitial(IntitialNode)
createActivityFinal(ActivityFinalNode)
createFlowFinal (FlowFinalNode)
createPaths(ActivityR)
createPath(start, finish, isJoin, guard)
createTokenEngines(ActivityR)
createTokenEngine(Queue, engineType)
createJoinCriteria(PullEngine)
```

**Fig. 4** ActivityFactory class and its operations

#### B. Activity Construction

The following code shows process of the instance creation:

```
public ActivityR createActivity(Activity actD) {
 actR = ActivityR.addNew(); // create empty activity
 for (element in actD.activityNode) { // for each ActivityNode
   switch (oclIsTypeOf(element)) { case (InitialNode) {
     createInitial(element); // create Inital node
     createOutputQueue(element);} // and create queue for it
   ... /* similarly create runtime elements for
       CallBehaviorAction, InputPin, OutputPin, ForkNode,
       JoinNode, DecisionNode, MergeNode */
   case (FlowFinalNode) {
    createFlowFinal(element);
         // create FlowFinal or node and
    creteInputQueue(element); } // create queue for it
   ... /* similarly for ActivityFinalNode */
```

```
     case (ActivityParameterNode) continue; }
         // skip ActivityParameterNodes
  for (element in actD.formalParameter) {
    creteOutputQueue(element); // create OutputQueues
    actR.parameter.addNew(element.parameter);}
         // and input parameters
   . . . /* similarly for returned parameters */
  for (element in actD.edge)
    createEdge(element);    // create edges
  createPaths(actR);             // create paths
  createTokenEngines(actR);   // create TokenEngines
  if (actR.finalNode->select(oclIsTypeOf(ActivityFinal))-
   >isEmpty()) actR.isFinal = False; // check for Final nodes
  actR.activate(); // activate the Activity
  return actR; }; // and return reference to this activity
```

The returned reference to the activity runtime instance is used for further activity management.

*C. Detailed View of create…() Methods*

Each *create..()* method consists of several steps. At first, relevant definition instances are obtained via definition associations, then already existing runtime instances are got, then for the current definition its runtime instance is created and linked with existing runtime instances, and definition-runtime associations are set. Due to size limitations, only the body for *createEdge(ObjectFlow edgeD)* method is shown:

```
public void createEdge(ActivityEdge edgeD) {
  edgeR = Edge.addNew(); // create new runtime Edge
  if (edgeD.source.oclIsKindOf(ObjectNode))
    edgeR.source = edgeD.source.runtime;
     // if source is ObjectNode, set proper OutputQueue
  if (edgeD.source.oclIsKindOf(ControlNode) and not
    edgeD.source.oclIsTypeOf(InitialNode))
    edgeR.source = edgeD.source.runtime;
    // if source is ControlNode, set proper IntermediateNode
  if (edgeD.source.oclIsTypeOf(InitialNode))
    edgeR.source = edgeD.source.runtime.outputQueue;
    // if source is InitialNode, set proper OutputQueue of the
      runtime Initial node
   . . . /* similarly set associations for targets */
  edgeD.runtime = edgeR;
   // update runtime association for definition element
  edgeR.definition = edgeD; };
   // update definition association for runtime element
```

*D. Path Construction*

Each activity diagram is a directed graph, thus we can use **path** with the same semantics as in the graph theory. In our case a path is a "transitive closure" of Edges and IntermediateNodes between Queues of StableNodes. This is realized in the *createPaths()* method. In our approach all constraints and conditions, coming from the relevant ActivityEdge Guards and ControlNodes are "concatenated" into the PassRule for each Path (e.g., order=approved AND sum>100). If a path has Joins, its attribute hasJoin is set to True, else it is False. The hasJoin attributes will be used for attaching paths to appropriate token engines.

The following code shows, how paths are created for the ADVM. The traditional "wave-front" algorithm for graph processing is used:

```
public void createPaths(ActivityR actR) {
  for (snode in actR.stableNode) { // for each StableNode
    for (oque in snode.outputQueue) { // for each OutputQueue
      inodes[] = null; inodescond[] = null; // reset
      start = oque; // start for path
      edges[0] = start.outgoing; // get 1st outgoing edge
      for (i <= inodes.length) {  // for unprocessed inodes
        for (edge in edges) { // for unprocessed edges
          guard = edge.guard.expression + " AND " +
           inodescond[i - 1]; // concatenate conditions from edge
           and outgoing inode (for brevity -  empty element (or
           outside array) is True)
          if (edge.target.oclIsTypeOf(InputQueue))
             // get target, if it is InputQueue,
            createPath(start, edge.target, isJoin, guard);
             // create new path
          else { // if it is intermediate node
           if (edge.targetNode.oclIsTypeOf(Join))
             isJoin = True; //  if path has join
             // mark it, don't need more check for valid ADs
           inodes[j] = edge.targetNode;
             // store the next unprocessed inode
           inodescond[j] = guard; // store "preconditions"
           j++;}} // count next unprocessed inode
        edges = inodes[i].outgoing; // get edges from inode
        i++;}}}};// count next processed inode
```

*E. TokenEngines Construction*

The *createTokenEngines()* method creates TokenEngines for queues in VM in the following way:

- if for an OutputQueue there exists at least one outgoingPath with hasJoin=False, then a PushEngine for this OutputQueue is created.
- if for an InputQueue there exists at least one incomingPath with hasJoin=True, then a PullEngine for this InputQueue is created.

The same restrictions for OutputQueue in OCL look the following way:

```
context OutputQueue
inv: OutputQueue.output->exists(hasJoin=False)
 implies
 OutputQueue.engine->one(oclIsTypeOf(PushEngine)) and
 OutputQueue.pushEngine.path-
  >exists(oclIsTypeOf(PushPath)) and
 OutputQueue.pushEngine.path-
  >select(oclIsTypeOf(PullPath))->isEmpty()
```

The similar restrictions apply to InputQueues. The rules imply the following consequences:

- PushEngines process only PushPaths, but PullEngines – only PullPaths.
- Queues can be without TokenEngines and paths of a Queue can be processed by several TokenEngines.
- Paths can have TokenEngines at both ends, but then these paths are processed by only one TokenEngine (either PushEngine or PullEngine).

Construction of PushEngines is simple, because for them all paths are independent (because of mutually exclusive guards for edges). Creation of PullEngines is more complicated, because for PullEngines additional conditions for path joinCriteria are necessary. The following code shows, how TokenEngines are created:

```
public void createTokenEngines(ActivityR actR) {
  for (snode in actR.stableNode) { // for each StableNode
    . . . /*create PushEngines, and continue with PullEngines */
    for (ique in snode.inputQueue) { // for each InputQueue
      for (path in ique.input) { // for each incoming Path
        if (path.isJoin == True) { // if some path has Joins
          if (ique.pullEngine.oclIsEmpty())
            engine=createTokenEngine(ique,"PullEngine");
              // create new PullEngine, if doesn't exist yet
          engine.pullPath.addNew(path); }}
              // add PullPath to PullEngine
      createJoinCriteria(engine); }}}
        // create joinCriterion for PullPaths of this PullEngine
```

Fig. 5 shows significant cases, how token engines are created and how they are linked to queues and paths. If a queue has an engine, it is filled with color, and "Push" or "Pull" beside it means the engine type. Paths, which are processed by an engine, are shown in the engine's color. If some edge is part of several paths, it is replicated in several colors.

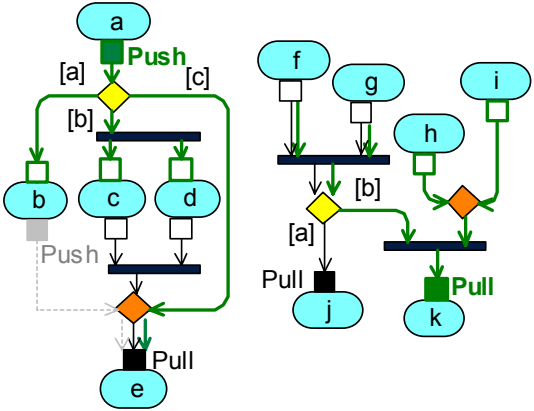

**Fig. 5.** Examples of activities with Queues, Paths, PushEngines and PullEngines, and their relationship

*F. Creation of Join Criteria for Pull Paths*

As it was mentioned before, for PullPaths having a common target an additional joinCriterion must be created. The JoinCriterion has to check whether a group of supplied tokens (one from each selected OutputQueue, in a subset of all potentially available queues) can be jointly moved to the InputQueue according to AD semantics. The JoinCriterion is determined by PullPaths and their nodes, and the sequence of nodes determines the structure of the expression. If paths have a common Join node, then tokens from all incoming paths should be joined (AND), if paths have a common Merge, then token coming from any one path can be used (OR), and if paths have a common Decision node, it can be ignored, because it doesn't play a role in token joining. Thus the JoinCriterion is a boolean expression, which is obtained by going upstream from an InputQueue serviced by a PullEngine and adding AND operation for a Join, with incoming edges as its operands (similarly, OR for Merge), until we reach the output queues, which play the role of elementary variables in this expression. Certainly, only edges of PullPaths are used in this process. It can be easily seen that the fan of incoming PullPaths (presented in the natural way with common segments overlapping), in fact, is equivalent to a tree form of the corresponding JoinCriterion.

Guard conditions for edges have already been included into passRules for paths, therefore only joinSpecifications of Join nodes must be added to joinCriteria.

The createJoinCriteria(PullEngine) method creates JoinCriterion for an engine. It uses the "wave-front" algorithm (in a way similar to IV.D), but going upstream from the PullEngine's InputQueue, over all PullPaths. When the tree form of the expression has been built, it is "translated" by traditional methods into a textual prefix form (which is used for evaluation on token sets during runtime). While there are a lot of technical steps in all this, the general idea is simple enough, so we do not provide more details of the method.

From an example in Fig. 6 the following expression is created:

OR(AND("p1.att2 = p2.att2", p1, p2), AND(p2, p3))     (1)

where OR is for the Merge node, and two ANDs are for Join nodes. This is a "shorthand notation" for the expression, its semantics is defined by rewriting it to a detailed OCL constraint (also in the prefix form!):

or (and (tokens->select(locus=p1).att2 = tokens->select(locus=p2).att2, tokens->exists(locus=p1), tokens->exists(locus=p2)), and (tokens->exists(locus=p2), tokens->exists(locus=p3)))     (2)

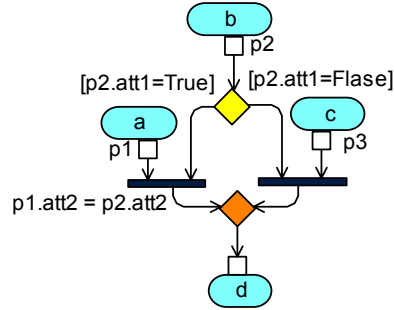

**Fig. 6.** Examples for joinCriteria construction for PullPaths

Here p1, p2, p3 are names of output queues (pins), which in this case represent the pin type. Use of pin names is convenient, because normally each pin has a type in our subset. Uniquely typed pins are used only for the example simplicity, a similar expression could be built, using pin names, but not types (unique pin names could be generated automatically during the runtime creation, if types are not unique within a "pull region").

The expression (2) is set as PullEngine.joinCriteria.Expression for the PullEngine of the action *d*. The use of JoinCriteria will be described in details in section V.D.

## V. EXECUTION AND DETAILED SEMANTIC OF VIRTUAL MACHINE CLASSES

Fig. 7 shows the metamodel of our ADVM. This diagram is another view of the same metamodel, which was shown in Fig. 3, but with more detailed runtime classes. It shows the complete set of classes, associations and operations, which are necessary for execution of ADVM. Runtime behavior of main classes will be described in details in the following sections.

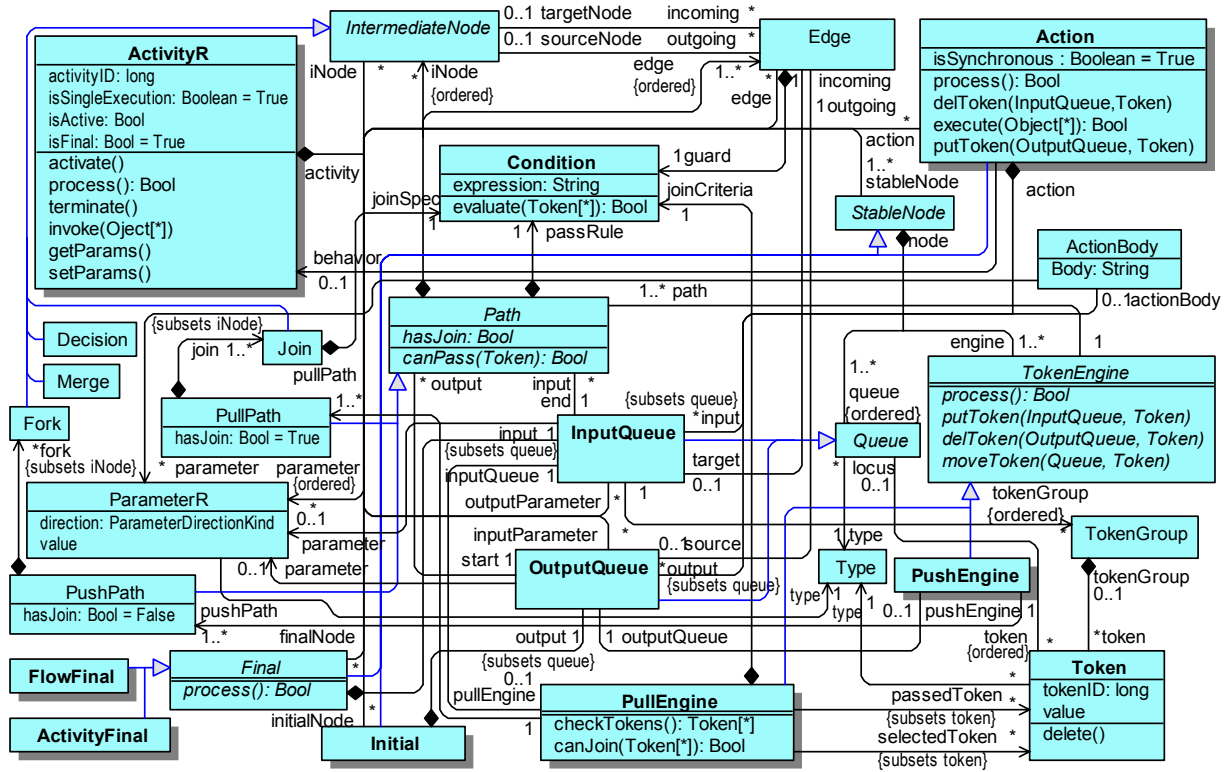

**Fig. 7.** Metamodel of the Activity Diagram Virtual Machine

### A. Activity Starting

Activity activation is done through asynchronous invocation of *process()* operations for all stable nodes and the activity, and these processes are running while the isActive flag for the activity is True. This operation is described using OCL:

**context ActivityR::activate()**
**post**:
  ActivityR.isActive = True and
  -- activity is active and all process() methods are activated
  ActivityR.process() and
  ActivityR.stableNode->select(oclIsTypeOf(Action) Or
  oclIsKindOf(Final))->forAll(n | n.process())

The *invoke()* operation starts an activity using actual input arguments, which are passed by the caller (see V.E) and returns actual output parameters from this activity. *invoke()* works with generic data and knows nothing about tokens. We assume here that actual and formal input parameters (and also the relevant parameter nodes) are ordered the same way. The method is described by a pseudocode:

```
public void invoke (Object[] data) {
 params =  ActivityR.parameter->select(direction = "in");
 for (i < data.length) {
    params[i].value = data[i]; i++;}
            // set actual argument values in input parameters
 setParams(); // transfer passed parameters to activity nodes
 getParams(); // wait for results and get from activity nodes
 params = ActivityR.parameter->select(direction = "out");
 for (i < params.length) {
    data.addNew(params[i]);i++;}} // set output values at the
end of input data
```

The *setParams()* method passes activity input parameter values as new data tokens in appropriate input Parameter nodes. It is described by OCL:

**context ActivityR::setParams()**
**post**:
  **let**
    input : OrderedSet(ParameterR) = ActivityR.parameter->select(direction = "in"),
  -- set of parameters
    cTokens : Set(Token) = ActivityR.initialNode.output.token**@pre**,
  -- existing control tokens in initial nodes
    dTokens : Set(Token) = ActivityR.inputParameter.token**@pre**
  -- existing data tokens in parameterNodes
  **in**
  **if** input->isEmpty() and ActivityR.initialNode->notEmpty()
  **then**
  -- if activity is invoked without params, use initial nodes
    ActivityR.initialNode->forall(n |
  -- set tokens for all Initial nodes
      (n.output.token-cTokens).oclIsNew() and

  -- new Token is created in OutputQueue of Initial node
      (n.output.token-cTokens).oclIsTypeOf(Null))
-- and its type is Null (control token)
  **else**
    input->forAll(ip | ActivityR.inputParameter-
>select(parameter = ip).oclIsNew()) and
-- for each parameter, its parameterNode has new token
      (ActivityR.inputParameter.token-dTokens).value =
input.value -- values from parameters are set to data tokens,
both ordered sets must be equal - elements are equally ordered
  **endif**

*getParams()* method is similar to the *setParams()* method with the only difference that it selects data tokens from outputParameter nodes of the activity and places values into output parameters, therefore it will not be described in detail.

The *process()* method for an activity is running, if the activity has no final nodes. It works similarly as for a final node and stops the activity when all its outputParameter nodes are filled with tokens:

**context** ActivityR::process()
**pre:**
  ActivityR.isActive and ActivityR.isFinal = False and
-- activity is active and has no ActivityFinal nodes
  ActivityR.outputParameter->forAll(token->notEmpty())
-- and all outputParameter nodes are filled
**post**:
  ActivityR.isActive = False -- activity is finished

The *terminate()* method is necessary for termination of subordinate activities. It terminates an activity without any conditions.

**context** ActivityR::terminate()
**post:**
  ActivityR.isActive = False and -- activity is stopped and
  ActivityR.stableNode.queue.token->isEmpty();
-- all tokens are deleted

*B. TokenEngines*

TokenEngines are working continuously and they move tokens from OutputQueues to InputQueues.

Each time when a new token appears, the evaluation of *canPass(Token)* is performed for this token and all relevant paths. For PullEngines additionally *canJoin(Token[*])* is reevaluated for all token subsets which can be formed from those located in relevant OutputQueues.

*C. Push Engine*

For Push engines, each token can be processed apart from any other token, and results depend on nothing more than the token itself. If the token passes the passRule of at least one PushPath, it is put on all the appropriate InputQueues immediately and removed from the OutputQueue. Because our model doesn't support global activity parameters, if a token can pass no PushPaths, it doesn't need to be checked again and sticks in the OutputQueue forever, if there are no any other paths and pull engines which can process this OutputQueue.

The following OCL code shows work of the PushEngine:
**context** PushEngine::process()
**pre**:
  PushEngine.action.activity.isActive and  -- Activity is active
  PushEngine.outputQueue.token->select(t |
t.locus.output.canPass(t))->notEmpty()
-- token is in OutputQueue, which can pass at least one Path
**post**:
  **let** prevToken : Token =
PushEngine.outputQueue.token**@pre**-
>select(token.locus.output.canPass(token))->first() **in**
      -- get 1st token from available
  **let** queues : Set(InputQueue) =
PushEngine.outputQueue.output-
>select(canPass(prevToken)).end **in**
      -- get set of InputQueues, for "passed Paths"
  **let** existTokens : Set(Token) = queues.token**@pre in** -- get set of tokens, which were in available InputQueues before
    PushEngine.outputQueue.token->select(token =
prevToken)->isEmpty() and
  -- now this token doesn't exist in OutputQueue and
    (queues.token-existTokens)->forAll(t | t.oclIsNew()) and
  -- in available InputQueues are new tokens
    (queues.token-existTokens)->forAll(token.value =
prevToken.value)
  -- and each new token has the same value as previous

*D. Pull Engine*

PullEngines exhibit the complicated semantic of token management. Similarly to the PushEngine, each token from each OutputQueue should be validated against the passRule of the relevant PullPath. If the token passes the passRule, it is marked as passedToken and can be processed further. If the token passes no passRule it doesn't need to be checked again and will stick in the OutputQueue forever, if there are no any other paths and engines which can process this OutputQueue.

In contrast with PushEngine, a PullEngine has more than one OutputQueue and any token should be moved in dependence from other tokens in respective OutputQueues.

Each time, when a new token appears in some OutputQueue, the engine's *process()* operation checks this token using the *canPass()* method for its PullPath, and, if the token passes, it is added to passedToken list. Then the *process()* operation invokes the *checkTokens()* method, which scans all subsets of the passedTokens set and tests them against the relevant JoinCriterion (by means of the *canJoin()* method, which evaluates the criterion). As soon as a valid subset is found, it is stored in the selectedToken list and the scan is terminated. Then the *process()* operation joins the newly found set into the InputQueue. Tokens are joined in the following way:

- If all tokens are control tokens, then one control token is posted in the InputQueue.
- If some of the tokens are data tokens, then all data tokens are posted in the InputQueue and they are grouped in a new TokenGroup.

To explain better the general principles of PullEngine behavior, the same operations are described in both ways – using pseudocode and OCL expressions: The following pseudocode shows the dynamic sequence of method invocation for PullEngine:

**public class PullEngine** {
**public process**() {
  **while** (PullEngine.activity.isActive) {

```
    if (PullEngine.inputQueue.input.start.token->exists(t |
t.oclIsNew())) { // if new token is in any OutputQueue
      thisToken = PullEngine.inputQueue.input.start.token-
>select(t | t.oclIsNew())->first(); // take it and
      if (PullEngine.PullPath.canPass(thisToken)) {
            // check it against passRule
        passedToken.addNew(thisToken);
            // if passed, update passedToken list
      if (checkTokens() <> Null) {
        if (selectedToken->forAll(type=Null)) {
          PullEngine.putToken(InputQueue,Null);
            // if all are control tokens, create on control token
          (for token in selectedToken)
PullEngine.delToken(token.locus, token);}
 // and delete processed control tokens form OutputQueues
        else { // if data tokens
          (for token in selectedToken->select(type<>Null)) {
            group=TokenGroup.addNew();
              // create new token group
            PullEngine.moveToken(inputQueue, token);
              // move all data tokens to InputQueue
            token.group = group;}}}}}}
              // and add them to the same group

public Token[] checkTokens() {
  while (selectedToken.length > 0) {
// while a new subset of passedToken set exists
    selectedToken = nextComb(passedToken); // get next
    if (canJoin(selectedToken)) // and check for joinCriteria
      return selectedToken}} // if can join, return

public Boolean canJoin (Token[] selectedToken) {
  return PullEngine.joinCriteria.evaluate(selectedToken) {
  // uses PullEngine.joinCriteria.Expression and
selectedToken,
  }}} // and returns True or False

public class Condition {
public Boolean evaluate (Token[] tokens) {
// returns evaluation of expression, where tokens are
referenced as variables (by pin names)
  return eval(expression, tokens);}}
```

The following OCL expression shows the token movement principles for PullEngine:
**context PullEngine::process()**
**pre**:
  PullEngine.action.activity.isActive and -- Activity is active
  PullEngine.checkTokens()->notEmpty()
        -- there are tokens which can be joined
**post**:
  **let** ique = PullEngine.inputQueue,  -- engine's InputQueue
    prevTokens : OrderedSet(Token) =
PullEngine.checkTokens@pre (),  -- processed tokens
    existTokens: OrderedSet(Token) =
PullEngine.inputQueue.token@pre, -- other existing tokens
    existGroups: OrderedSet(TokenGroup) =
PullEngine.inputQueue.tokenGroup@pre
-- other existing TokenGroups in InputQueue
  **in**
    (ique.input.start.token-prevTokens)->isEmpty() and
-- now these tokens are removed from OutputQueues
  **if** prevTokens->reject(oclIsTypeOf(Null))->notEmpty()
  **then**     -- if there were object tokens
    (ique.tokenGroup-existGroups).oclIsNew() and
          -- new tokenGroup is created
    (ique.token-existTokens)->forAll(t | t.oclIsNew()) and
          -- new object tokens exist in InputQueue
    (ique.token-existTokens) = prevTokens-
     >reject(oclIsTypeOf(Null)) and
    -- with data from prevTokens (comparing as ordered sets)
    (ique.token-existTokens).tokenGroup = ique.tokenGroup-
existGroups -- and they are included in the same new group
    **else**  -- if all were control tokens
    (ique.token-existTokens).oclIsNew() and
        -- one control token is created in the InputQueue
    (ique.token-existTokens).oclIsTypeOf(Null)
  **endif**

*E. Action*

Actions have processes running all the time (for checking tokens in their InputQueues) and they consume tokens from their InputQueues and provide tokens in OutputQueues. Tokens are checked again, when a new token appears in any InputQueue. An Action will only start execution, if all its InputQueues are filled.

Tokens from InputQueues are consumed when all InputQueues have at least one token (an Action works as an implicit join). If a token has no TokenGroup, one token from this InputQueue is consumed; else all tokens from this group are consumed. The consumption means that the action engine extracts data from data tokens and stores these data as the actual argument list for invocation. Then it executes the ActionBody using *execute(Object[*])*, if it is an opaque behavior, or creates and invokes another activity using the *createActivity()* and *invoke(Object [*])* operations. If isSynchronous = True, the action waits for output, and the actual results are placed as data tokens into OutputQueues. Else, control tokens are placed into OutputQueues immediately after the execution/invocation. The following pseudocode shows the behavior of the action:

```
public class Action {
public process() {
  while (Action.activity.isActive) { // activity is active
    if (Action.input->forAll(token->notEmpty())) {
          // -- for each InputQueue exists a token
      for (token in Action.input) {
        if (token.tokenGroup->notEmpty()) {//if tokens are
grouped in a group
          for (gtoken in token.tokenGroup) // get them from the
group
            intokens.addNew(gtoken); // store them in a list
        else intokens.addNew(token); // else get only this token
        indata = intokens->select(type<>Null).value; // get data
from all data tokens
      if (Action.isSynchronous = True) { // if synchronous
        if (Action.behavior->notEmpty()) { // if it invokes
another activity
          activity =
(ActivityFactory.createActivity(Action.behavior));
```

```
            // create the activity runtime
        activity.invoke(indata); // invoke the activity
/* we assume order of pins conform to order of parameters */
        outdata = activity.parameter->select(direction =
"out").value;}  // get returned arguments, which are stored in
output parameters
     else { // if it is opaque behavior
        action.execute(indata) // execute it
        outdata = Action.ActionBody.parameters-
>select(direction = "out").value;}  // and get returned
arguments
     for (i < outdata.len) {
     token.addNew(locus = Action.output[i], type =
outdata.type, value = outdata);}};
// set tokens in OutputQueues with right type and value
     else { // if invocation is asynchronous
       if (Action.behavior->notEmpty()) // if refers to activity
(AcytivityFactory.createActivity(Action.behavior)).invoke(in
data); // invoke it asynchronously
       else action.execute(indata); // or action asynchronously
     for (output in Action.output) // put control tokens
       putToken(output,Null);}}}}}} // in all OutputQueues
```

*F. Activity Finishing*

Final nodes also have running processes, which process tokens. FinalNodes simply delete tokens from their inputQueues:

**context FlowFinal::process()**
**pre**:
  FlowFinal.activity.isActive and  -- Activity is active and
  FlowFinal.input.token->notEmpty()  -- at least one token is in input queue
**post**:
  FlowFinal.input.token->isEmpty() -- this token is deleted

If a token reaches an ActivityFinal node, all tokens, excluding those which are in output parameters of the activity are deleted and all actions are stopped; invoked activities are stopped and their tokens are deleted without any conditions.

```
public process() {
  while (ActivityFinal.activity.isActive) {  // activity is active
   if (ActivityFinal.input.token->notEmpty()) {
     ActivityFinal.activity.isActive  = False; // stop activity
    for (token in (stableNode.queue -
ActivityFinal.activity.outputParameter).token)
       token.delete(); // delete tokens except in outputParameter
nodes
     activities[0] = ActivityFinal.activity;
        // get first activity
     for (i <=activities.length) { // for each invoked activity
       action = activities[i].action; // get action in next activity
       for (action in actions)
         if (action.behavior->notEmpty()) // if action invoked
             another activity
           activities.addNew(action.behavior); // add it to list
       activities[i].terminate;  // terminate current activity
       i++; }}// go to next activity
```

As it was mentioned before in section V.A, if the activity has no final nodes, it can also be finished, when all its outputParameter nodes are filled.

## VI. PROOF OF ADVM EQUIVALENCE

In this section we provide a semiformal proof of the equivalence of the original semantics (ADVM) of UML activity diagrams [1] with the one provided in this paper. We remind that the original semantics is based on token offering (visibility), control nodes acting as distributed switches and actions pulling tokens "all at once" (see III A).

We **assert** that the **essential event trace** – starts of action executions and the token sets consumed and produced by these actions **are the same for both virtual machines on any activity diagram** in the subset described in section II. The events in this event trace occur in the same order and in the same time moments. Moreover, each individual token traverses the same path according to both machines, but the ordering of these token movements in time may differ. In general, in our VM tokens will reach their destinations earlier than in the original VM.

At first, we will show that there is no real race for tokens by actions in the selected subset. More precisely, no token can be potentially delivered via several alternative paths by several token engines. Certainly, such conditions cannot appear for push paths – because using push engines we deliver each token along all of the paths from an output queue, where guards permit. These could be only pull paths where (forward) branch points occur (merge points backward from the pull engine prospective).

We prove that no token in an output queue may be serviced by (be in the valid token set for) more than one pull engine (but the same output queue may be).

Let us analyze where the (forward) branch points in a path may occur:

- If multiple edges leave the same action or object node (implicit fork). In our subset it is not allowed, each output pin has exactly one outgoing edge and each edge leaving an action starts from an output pin. Object nodes and central buffer nodes, which allow competing outgoing edges, are not included in our subset.
- When a fork appears in some path. But there may be no forks in a path leading to a pull engine in our subset (because there must be an explicit join in a pull path).
- When Decision allows a token to traverse through several edges. In our subset any token can traverse only one edge (because of mutually exclusive guards, only one guard is true). So, though several pull engines may have a common source (output queue), for each token only one path is enabled – that where the guards on the path to it are true.

It has to be proven also, that there may be no races between push and pull paths. In other words, if a token is delivered by one or more push paths (and then removed from the source queue), it may not be "useful" for a pull path starting from the same queue. Indeed, it is so because the passRule may be true for only one path, if this path contains no forks, and pull paths do contain no forks. Similarly, if a token is "pulled" by a pull path, it is "of no interest" for any push path. Thus, our subset is race free.

Hence follows that each token can be delivered to only one destination (or several ones for push paths). The two preceding sections where our VM was described in detail actually provide an assurance that this target is the same as in the original activity diagram semantics.

Since or VM delivers tokens along push paths as soon they appear in an output queue, and each pull engine independently pulls a token (or a group of tokens) into its input pin as soon as they are available on relevant outputs, in general any token will be delivered to its destination in our VM not later than in the original VM. In fact, frequently it will be earlier, since the original VM always transports a group of tokens into all input pins of an action simultaneously.

It remains to show that our simplified action engine which takes tokens or groups of tokens directly from input pins, cannot be activated earlier than the original one which takes the tokens from output pins. Our rule asserting that all input pins must have at least one token (or an appropriate group of tokens, if this is a "join pin") for the action to start assures that actually it is exactly the same situation where the original VM would have finally collected all the offered tokens, which are required for the start of the action. The original principle of offering tokens via an edge downstream (as it was already noted in section III) actually means that tokens are made visible to the outgoing edge of a control node. Since finally this visibility is required for all input pins of the action, the moment for this action to fire is when the last token becomes available in the corresponding output pin. But it is exactly the same moment when our corresponding token engine (push or pull) would fill the last required input pin, and in the same moment our action engine would fire too.

This completes the proof of equivalence.

## VII. Conclusion

### A. Extensions and Practical Usage

The subset of the UML activity diagrams has been chosen to cover the basic needs for business process management and workflow systems. The goal was to show that AD notation can be given a precise enough "natural" semantics, so that a workflow engine can be based on it. From the graphical syntax point of view, according to [2] AD is an acceptable and at the same time well-known notation for the workflow definition. Another usage of ADVM could be for an AD simulator – a tool important for workflow validation.

Certainly, some elements from the Complete (or even Complete structured) level are of high value for workflow definitions. Many of these features actually could be dealt with by the methodology proposed in this paper. Events and even interruptible regions could be treated much the same way. A more serious problem is the use of context attribute values in guards. Treating them as specified in [1] would lead to a continuous recheck, when a token is offered to an edge, but rejected by the guard. A solution is to assume that guards are evaluated for the current "snapshot" of the context and any context changes are modeled as explicit events. This solution could be easily implemented in our ADVM.

A technical issue is the exhaustive search (among all token subsets) for joinCriteria, which was used only for the simplicity of description, for real systems it can be made efficient by converting a join criterion to disjunctive normal form (DNF) and checking only tokens in each AND-term. A CentralBuffer (DataStore) with one outgoing edge can be included as another type of StableNodes without model changes. If several outgoing edges and concurrency is necessary, a more complicated approach is required, but this case is not typical for workflow definitions.